\documentclass[twocolumn, switch]{article} 

\usepackage{preprint}

\usepackage{amsmath, amsthm, amssymb, amsfonts}

\usepackage[numbers,square]{natbib}
\usepackage[utf8]{inputenc} 
\usepackage[T1]{fontenc}  
\usepackage{xcolor}   
\usepackage[colorlinks = true,
            linkcolor = purple,
            urlcolor  = blue,
            citecolor = cyan,
            anchorcolor = black]{hyperref}  
\usepackage{booktabs}     
\usepackage{nicefrac}   
\usepackage{microtype}    
\usepackage{lineno}   
\usepackage{float}      

\usepackage{lipsum}   

\usepackage{newfloat}
\DeclareFloatingEnvironment[name={Supplementary Figure}]{suppfigure}
\usepackage{sidecap}
\sidecaptionvpos{figure}{c}

\usepackage{titlesec}
\titlespacing\section{0pt}{12pt plus 3pt minus 3pt}{1pt plus 1pt minus 1pt}
\titlespacing\subsection{0pt}{10pt plus 3pt minus 3pt}{1pt plus 1pt minus 1pt}
\titlespacing\subsubsection{0pt}{8pt plus 3pt minus 3pt}{1pt plus 1pt minus 1pt}

\usepackage{tikz,xcolor,hyperref}

\usepackage{graphicx}
\usepackage{caption}
\usepackage{subcaption}
\usepackage{stfloats}
\usepackage{amssymb,amsmath,amsthm,mathtools}
\usepackage{bbm,physics,braket,mathrsfs}

\definecolor{lime}{HTML}{A6CE39}
\DeclareRobustCommand{\orcidicon}{
  \begin{tikzpicture}
  \draw[lime, fill=lime] (0,0) 
  circle [radius=0.16] 
  node[white] {{\fontfamily{qag}\selectfont \tiny ID}};
  \draw[white, fill=white] (-0.0625,0.095) 
  circle [radius=0.007];
  \end{tikzpicture}
  \hspace{-2mm}
}
\foreach \x in {A, ..., Z}{\expandafter\xdef\csname orcid\x\endcsname{\noexpand\href{https://orcid.org/\csname orcidauthor\x\endcsname}
      {\noexpand\orcidicon}}
}

\title{Analysis of Quantum Steering Measures}

\usepackage{authblk}

\author{ \large Lucas Maquedano\orcidA{}}
\author{Ana C. S. Costa\orcidB{}}

\affil{Department of Physics, Federal University of Paran\'a, Curitiba, PR, Brazil}

\begin{document}

\twocolumn[ 
  \begin{@twocolumnfalse} 
  
\maketitle

\begin{abstract}
The effect of quantum steering describes a possible action at a distance via local measurements. In the last few years, several criteria have been proposed to detect this type of correlation in quantum systems. However, there are few approaches presented in order to measure the degree of steerability of a given system. In this work, we are interested in investigating possible ways to quantify quantum steering, where we based our analysis on different criteria presented in the literature.
\end{abstract}
\vspace{0.35cm}

  \end{@twocolumnfalse} 
] 



\section{Introduction}
\label{Introduction}

In 1935, Einstein, Podolsky, and Rosen~\cite{Einstein1935} pointed out contradictions between local realism and the completeness assumptions of quantum mechanics. In response to this argument, Schrödinger~\cite{Schroedinger1935I, Schroedinger1935II, Schroedinger1936} introduced the term {\it{steering}} to present the idea that the source of these contradictions was based on Alice's ability, from her choice of measurement basis, to influence Bob's state. This debate gave rise to a new field in physics, dedicated to exploring these quantum correlations.

The study of quantum correlations has contributed fundamentally to the area of quantum information and various aspects of the foundations of quantum theory. From the point of view of applications, quantum correlations have been used for the development of several quantum information and quantum computing protocols. Unlike Bell's nonlocality~\cite{Brunner2014} and entanglement~\cite{Horodecki2009}, {steering} 
\cite{Wiseman2007} is fundamentally asymmetrical. It allows, for example, entangled systems that are only steerable from one observer to another, known as one-way steering~\cite{Bowles2014,Skrzypczyk2014,Zeng2022,Sekatski2023}. This property makes quantum steering an interesting resource for quantum information processing where some of the parties are considered untrusted, such as quantum key distribution~\cite{Branciard2012,Lo2014}, randomness generation~\cite{Law2014,Passaro2015,Skrzypczyk2018}, subchannel discrimination~\cite{Piani2015}, and quantum teleportation~\cite{Reid2013}.

Quantum systems may describe different types of correlation, including nonlocality~\cite{Brunner2014}, steering~\cite{Reid1989,Cavalcanti2009,Cavalcanti2017,Uola2020}, and entanglement~\cite{Horodecki2009,Guhne2009}.  There is a hierarchy among them, meaning that quantum states that present Bell's nonlocality are also steerable and entangled, but not all entangled states are steerable and violate a Bell inequality~\cite{Wiseman2007,Quintino2015}. In a practical scenario, trust in the measurement apparatus plays a role here, where certifying entanglement requires trusted devices on both sides of the experiment, requires that Bell's nonlocality is demonstrated with untrusted devices on both sides, and that steering is possible in a one-trusted device~scenario.

Quantum steering is a well-defined type of quantum correlation, and there are several ways to detect it; however, its quantification is still a matter of debate. Several proposals have been presented in the literature, where we have, for example, the steering weight~\cite{Skrzypczyk2014} and the steering robustness~\cite{Piani2015}. Another attempt in the direction of quantifying quantum steering was given in~\cite{Costa2016}, where the authors based their measure on the maximal violation of a linear steering inequality~\cite{Cavalcanti2009}.

In this work, we are interested in analyzing different steering quantifiers. Our goal is to compare the existing measures of quantum steering with new proposals that we introduce here. Note that it is not our intention to declare that one measure is better than the other but to show that some are more efficient in a given context. We analyze the following criteria: the linear steering~\cite{Cavalcanti2009,Costa2016}, the generalized entropic steering~\cite{Costa2018,Costa2018-2}, the rotationally invariant steering~\cite{Wollmann2018}, and the dimension-bounded steering~\cite{Moroder2016}.

To do this, we associate each criterion with a quantifier based on the method presented in~\cite{Costa2016}, giving a steerability degree for a given state. Once we have the most advantageous quantifier, we compare it to a numerical measure based on semidefinite programming~\cite{Cavalcanti2017}. Finally, we consider the approach defined in~\cite{Fonseca2015}, where the authors considered the volume of violations of a given Bell inequality in order to quantify the degree of nonlocality in the system. We extend their analysis to quantum steering and compare it with our previous method for a specific example, considering a family of Werner states~\cite{Werner1989}.

\section{Quantum Steering}

In a typical quantum steering scenario, we consider two parties, Alice and Bob, who share a bipartite quantum state. In each round of the experiment, Alice performs different measurement settings $x$ with outcome $a$, while Bob remains with an unnormalized conditional state $\sigma^B_{a|x}$ for each of Alice's choices. The collection of these conditional states is called the {\sl {steering assemblage}
}. According to quantum theory, the members of the assemblage can be calculated by $\sigma^B_{a|x}:=\Tr_A(M^A_{a|x}\otimes \mathbbm{1}^B \rho)$, where $\{M^A_{a|x}\}_a$ is a set of positive operator-valued measure (POVM), with $\sum_a M^A_{a|x} = \mathbbm{1}$ and $M^A_{a|x}\geq 0$. Note that the conditional states must obey the condition $\sum_a\sigma^B_{a|x} = \rho^B$, which means that Bob's reduced state $\rho^B = \Tr_A(\rho)$ is independent of Alice's measurement choices.

After characterizing the assemblage, Bob wants to check whether the state he shares with Alice is entangled. This test can be done by checking if Bob's assemblage satisfies a local hidden state (LHS) model~\cite{Wiseman2007}, which is constructed by considering that Bob's particle is in some hidden state $\rho^B_\lambda$ with probability $p(\lambda)$, parametrized by a hidden variable $\lambda$. The outcome of a measurement performed by Alice would only give him extra information about Bob's state probability. According to this scenario, we arrive at~\cite{Wiseman2007}
\begin{equation}\label{LHSassemblage}
\begin{aligned}
\sigma^B_{a|x} = \int d\lambda p(\lambda) p(a|x,\lambda) \rho^B_\lambda.
\end{aligned}
\end{equation}
Equation~\eqref{LHSassemblage} is known as an LHS model, and violating this model would mean that Bob's state is steerable. In this case, it is possible for Alice to steer Bob's state in some nonlocal way. It is important to note that, unlike entanglement and Bell's nonlocality, quantum steering is an asymmetric quantum feature, and there are states where Alice can steer Bob but not the other way around~\cite{Bowles2014,Skrzypczyk2014,Zeng2022,Sekatski2023}.

\section{Detection of Quantum Steering}
\label{ineq}

In order to verify if a given state presents quantum steering, one needs to check whether it is possible to construct an LHS model, assuming that Alice performed a set of measurements and Bob ended up with unnormalized conditional states, which can be determined using a full tomography on such states. However, depending on the complexity of some states, this is not an easy task. One possibility is to make use of semidefinite programming (SDP) in order to verify whether the state assemblage satisfies an LHS model. Depending on the system, this technique might be demanding, given that it needs a full tomography on Bob's conditional states and an optimization procedure that can become intrinsically hard to perform. Given this difficulty, quantum steering can be detected by the violation of certain inequalities. The construction of these inequalities is based on an LHS model, and its violation would mean that it is not possible to construct such a model.

Thus, most of these inequalities are built using a combination of mean values (or probability distributions of certain outcomes) of specific measurements performed by Alice and Bob. Using the correlations presented by these results, one can check whether the shared system is steerable without needing to reconstruct the assemblage completely. In the following, we are going to present some criteria used to detect steering in bipartite quantum systems.

\subsection{Linear Steering Criteria}

In~\cite{Cavalcanti2009}, a linear steering criterion was proposed, based on the uncertainty relation between the average values of maximally incompatible observables, and it is given by
\begin{equation} \label{CJWR}
F^{(m)}_{\mathrm{L}}(\rho,\nu)= \qty|\sum_{i=1}^m\langle A_i \otimes B_i\rangle|\leqslant \mathcal{B}^{(m)}_{\mathrm{L}}.
\end{equation}
where $A_{i}=\hat{u}_{i} \cdot \vec{\sigma}$ and $B_{i}=\hat{v}_{i} \cdot \vec{\sigma}$, with $\vec{\sigma}$ a vector composed of the Pauli matrices ({{from now on,} 
we use the operators $A_i$ and $B_i$ to denote projective measurements performed by Alice and Bob, respectively}). Here, $\mathcal{B}^{(m)}_{\mathrm{L}}=\sqrt{m}$, where $m$ is the number of measurements performed at each site. Note that $\hat{u}_{i} \in \mathbb{R}^{3}$ are unit vectors, and $\hat{v}_{i} \in$ $\mathbb{R}^{3}$ are orthonormal vectors. The mean value is given by $\langle A_{i} \otimes B_{i}\rangle=\operatorname{Tr}[(A_{i} \otimes B_{i}) \rho]$, and $m\in \{2,3\}$. The violation of~\eqref{CJWR} means that the state is steerable.

Later on, these criteria were extended to a more general form~\cite{Saunders2010},
\begin{equation}\label{Iln}
\left|\sum_{i=1}^m\langle A_i\otimes B_i\rangle\right| \leqslant \max_{\{a_i\}}\left[\lambda_{\max}\left(\sum_{i=1}^n a_i B_i\right)\right].
\end{equation}
Here, $a_i = \pm 1$ and $\lambda_{\max}(X)$ is the highest eigenvalue of $X$. If we consider Alice and Bob performing $m=\{2,3\}$ observables, with Bob's measurements being orthogonal, we recover the linear steering criteria presented in Equation~\eqref{CJWR}.

It is important to note that Equation~\eqref{CJWR} depends on the choice of the measurements performed by Alice and Bob. Therefore, for a given state, it can happen that the inequality is violated by one set of Alice and Bob's measurements, and not by another one. In order to emphasize the dependence on the measurements, we inserted the $\nu$ symbol in the definition of $F^{(m)}_{\mathrm{L}}$.

\subsection{Generalized Entropic Steering}

A different type of steering criteria is based on entropic uncertainty relations. The first proposal~\cite{Walborn2011} considered the Shannon entropy and continuous variables, and in~\cite{Schneeloch2013}, the authors extended the criteria for discrete variables. Later, a generalization of the criteria was proposed considering the Tsallis entropy~\cite{Costa2018, Costa2018-2} and the Rényi entropy~\cite{Krivachy2018}. In this work, we focus on the proposal based on the Tsallis entropy, since numerical investigations suggest that it is more sensitive to detect steering than the one based on the Shannon entropy~\cite{Costa2018} and it is equivalent to the one based on Rényi entropy~\cite{Krivachy2018} for a given choice of parameters~\cite{Wollmann2020}.

Then, considering the Tsallis entropy~\cite{Tsallis1988} for a given probability distribution,
$S_q(\mathcal{P})=-\sum_n p^q_n \ln_q(p_n)$, with $\ln_q(p)=(p^{1-q}-1)/(1-q)$, the steering criteria derived in~\cite{Costa2018} are given by
\begin{equation}\label{tsc}
F^{(q,m)}_{\mathrm{E}}(\rho,\nu) = {-} \sum_{i=1}^m \left[S_q(B_i|A_i)+(1-q)C(A_i,B_i)\right] {\leqslant} \mathcal{B}^{(q,m)}_{\mathrm{E}},
\end{equation}
where $C(A_i,B_i)$ is a correction term, and $S_q(B_i|A_i)$ is the conditional Tsallis entropy~\cite{Furuichi2006}. For the case in which $q \in (0,2]$ and the measurements are mutually unbiased with dimension $d$, the bound in Equation~\eqref{tsc} can be calculated in the following way~\cite{Rastegin2013}
\begin{equation}
\mathcal{B}^{(q,m)}_{\mathrm{E}} = {-} m \ln_q \qty(\frac{m d}{m -1 + d}).
\end{equation}

It is also possible to rewrite this inequality in terms of the probabilities of Alice and Bob's measurements,
\begin{equation}
\label{tsc-prob}
F^{(q,m)}_{\mathrm{E}}(\rho,\nu) = {\frac{1}{1-q}}\left[\sum_{i=1}^m\left(1 - \sum_{ab}\frac{(p_{ab}^{(i)})^q}{(p_{a}^{(i)})^{q-1}}\right)\right] {\leqslant} \mathcal{B}^{(q,m)}_{\mathrm{E}}.
\end{equation}
Here, $p_{ab}^{(i)}$ is the probability of Alice and Bob for outcome $(a,b)$ when measuring $A_i\otimes B_i$, and $p_a^{(i)}$ are the marginal outcome probabilities of Alice's measurement~$A_i$, assuming that in this protocol, Alice aims to steer Bob's state. We are considering that Alice and Bob can perform $m$ measurements. The entropic criteria have been investigated in different contexts, and recently they were tested experimentally~\cite{Wollmann2020}.

\subsection{Rotationally Invariant Steering}

The rotationally invariant steering criterion, presented by Wollmann {et al.}~\cite{Wollmann2018}, is motivated by the fact that the steering can be verified, for a larger class of states, in a rotationally invariant way. Returning to the general scenario for steering presented at the beginning, after many runs of the protocol, Bob should be able to estimate the correlation matrix $\mathcal{M}_{ij}:=\left\langle A_{i} \otimes B_{j}\right\rangle$ and test whether it is compatible with an LHV model~\eqref{LHSassemblage}, such~that
\begin{equation}\label{MC}
\mathcal{M}_{ij}= \int d \lambda p(\lambda)\left\langle A_{i}\right\rangle_{\lambda}\left\langle B_{j}\right\rangle_{\rho^B_{\lambda}}.
\end{equation}
Restricting it to the case where all results are labeled by $\pm 1$ and Bob's measurements correspond to a set of orthogonal measurements, it was shown that any LHV model must satisfy the steering inequality
\begin{equation}\label{SRI}
F^{(m)}_{\mathrm{RI}}(\rho,\nu) = \|\mathcal{M}\|_{\mathrm{tr}} \leqslant \mathcal{B}^{(m)}_{\mathrm{RI}},
\end{equation}
with $\|\mathcal{M}\|_{\mathrm{tr}}:=\Tr\qty[\sqrt{\mathcal{M}^{\top} \mathcal{M}}]$ and $\mathcal{B}^{(m)}_{\mathrm{RI}} = \sqrt{m}$.

\subsection{Dimension Bounded Steering} 

Another criterion proposed in the literature is the dimension-bounded steering~\cite{Moroder2016}. It is based on the fact that an unsteerable assemblage can be reproduced by measurements in a separable state of the form
\begin{equation}
\rho=\sum_{a_1,\ldots, a_m}\ket{a_1,\ldots, a_m}_{A}\bra{a_1,\ldots, a_m} \otimes \omega^B_{a_1\ldots a_m},
\end{equation}
$x=\{1, \ldots, m\}$ being the number of measurements performed by Alice with $a_{x}$ outcomes. Here, the set of operators $\{\omega^B_{a_1\ldots a_m}\}$ is described by an LHS model. Since $\rho^B = \Tr_A(\rho)$ and $\rho^B = \sum_{a} \sigma^B_{a|x}$, it follows directly that an ensemble is \textit{unsteerable} if and only if there exist unnormalized density operators $\{\omega^B_{a_1\ldots a_m}\}$ such that the conditional states can be written as
\begin{equation}
\rho^B_{k | x}=\sum_{a_1, \ldots, a_{m}} \delta_{a_{x},k} \omega^B_{a_1\ldots a_{m}}.
\end{equation}
The authors follow with a method to remove the discord's zero structure from the state, exchanging $\ketbra{a_1\ldots a_m}$ for another operator of dimension $d_A$ and adding constriction relations in such a way that the resulting separable state $\Sigma$ becomes completely determined.

Considering dichotomic measurements and spin operators, the relevant criterion is evaluated through the data matrix $\mathcal{D}:=\Tr[(A_i \otimes B_j) \Sigma]$, with $A_0 = B_0 = \mathbbm{1}$. The determinant of the data matrix can be used to lower-bound the trace norm of a correlation matrix by adapting the CCNR criterion~\cite{Chen2003,Rudolph2005}, so one has a quantity for which an upper bound is known for separable states. This leads to the dimension-bounded steering inequality (for more details see~\cite{Moroder2016}):
\begin{equation}\label{F_DB}
F^{(m)}_{\mathrm{DB}}(\rho,\nu) = |\det \mathcal{D}| \leqslant \mathcal{B}^{(m)}_{\mathrm{DB}},
\end{equation}
with the bound
\begin{equation}
\mathcal{B}^{(m)}_{\mathrm{DB}} = \frac{1}{\sqrt{d_{A}}}\left(\frac{\sqrt{2 d_{A}}-1}{m \sqrt{d_{A}}}\right)^{m}.
\end{equation}
The violation of Equation~\eqref{F_DB} implies that the state is steerable. 

\section{Steering Quantifiers}
\label{measures}

So far, we have presented several steering criteria, given by the functions $F^{(m)}_{\mathrm{K}}(\rho,\nu)$ and their respective bounds $\mathcal{B}^{(m)}_{\mathrm{K}}$, with K = {L, E, RI, DB}. Although all of them can be used to detect steering in different contexts, and for some specific cases they all agree, these functions do not give the information of how steerable a given state is. Considering this fact, in this work, we are interested in proposing different ways to quantify the steerability of a given state, based on all functions presented above. Then, with these steering measures, we can compare them with themselves and with another well-known steering measure, the steering robustness~\cite{Piani2015}.

Our strategy is the following. We associate each inequality with a measure $S^{(m)}_{\mathrm{K}}$, which serves as a metric of the degree of steerability of a given state. Thus, $S^{(m)}_{\mathrm{K}} \in [0,1]$, meaning that this measure is one for maximally steerable states and zero for unsteerable states according to a given criterion ({{note that} 
the nonviolation of a given criterion does not mean that the system is unsteerable;~this answer is given only if it is possible to construct an LHS model for such a state considering infinite measurements}). In order to construct this measure, we consider the same approach used in~\cite{Costa2016}, where the authors proposed a steering quantifier based on the linear steering criteria~\eqref{CJWR}. In their work, they considered the amount by which a steering inequality was maximally violated. The reasoning behind this strategy is based on a similar approach used to quantify Bell's nonlocality in bipartite states~\cite{Brunner2014,Horodecki1995}. Operationally, one can think that a state that violates an inequality more is said to be more nonlocal (or quantum-correlated) because it is more resilient under the presence of noise.

Then, using the approach considered in~\cite{Costa2016}, the structure of the quantifiers associated with each inequality is given by
\begin{equation}\label{SLin}
S^{(m)}_{\mathrm{K}}(\rho):=\max \left[0, \frac{F^{(m)}_{\mathrm{K}}(\rho)-\mathcal{B}^{(m)}_{\mathrm{K}}}{F^{(m)}_{\mathrm{K}}(\rho_{\max})-\mathcal{B}^{(m)}_{\mathrm{K}}}\right],
\end{equation}
with $F^{(m)}_{\mathrm{K}}(\rho)=\max_{\nu} F^{(m)}_{\mathrm{K}}(\rho,\nu)$, meaning that we optimize over all measurements performed by Alice and Bob. In order to normalize this measure so it is restricted to the interval $[0,1]$, we consider the denominator of Equation~\eqref{SLin} such as $F^{(m)}_{\mathrm{K}}(\rho_{\max})$ to be the state that violates maximally the inequality. For bipartite systems, the Bell states, which are maximally entangled, are the ones that fulfill this feature.

Although we have written all $F^{(m)}_{\mathrm{K}}(\rho)$ considering $m$ measurements per site, we restrict our analysis to the case where Alice and Bob are performing three orthogonal measurements each. Moreover, we focus on general two-qubit states, which can be written in a Bloch representation~\cite{Luo2008} as
\begin{equation}\label{rho}
\varrho=\frac{1}{4}\left(\mathbbm{1} \otimes \mathbbm{1}+\vec{a} \cdot \vec{\sigma} \otimes \mathbbm{1}+\mathbbm{1} \otimes \vec{b} \cdot \vec{\sigma}+\sum_{r=1}^{3} c_{r} \sigma_{r} \otimes \sigma_{r}\right),
\end{equation}
where $\{\vec{a},\vec{b},\vec{c}\} \in \mathbb{R}^3$ and $\vec{a}^2 + \vec{b}^2 + \vec{c}^2 \leq 3$.

\subsection{Linear Steering Measure}

As we already mentioned, the quantifier for the linear steering criteria~\eqref{CJWR} was developed in~\cite{Costa2016}. Here, we reproduce some of their results because they are important in the analysis of other criteria we considered in this work.

Given that Alice and Bob perform projective measurements $A_i$ and $B_i$, respectively, we have that
\begin{equation}\label{ABmeanval}
\left\langle A_i \otimes B_i\right\rangle=\sum_{r=1}^{3} u_{i r} c_{r} v_{i r} \equiv\left\langle u_{i}|C| v_{i}\right\rangle \equiv C_{i},
\end{equation}
where $C \equiv \sum_{r} e_{r}\ketbra{e_{r}}$ is a Hermitian operator with eigenvalues $e_{r}$. Here, $\hat{u}_{i} \equiv \left|u_{i}\right\rangle=\sum_{r} u_{i r}\left|e_{r}\right\rangle$ and $\hat{v}_{i} \equiv\left|v_{i}\right\rangle=\sum_{r} v_{i r}\left|e_{r}\right\rangle$.

In~\cite{Costa2016}, the authors defined the transformation $ \left|\alpha_{i}\right\rangle \equiv C\left|u_{i}\right\rangle$, such that $C_{i}=\left\langle\alpha_{i} | v_{i}\right\rangle$, to maximize $\sum_i C_i$. The upper bound of this sum is obtained from the orthogonality between the vectors $\hat{v}_i$ and consequently, between $\ket{\alpha_i}$. Thus, the linear function (\ref{CJWR}) depends exclusively on the parameter $\vec{c}$ of the state $\varrho$, and
\begin{equation}\label{F_CJWR}
F^{(2)}_{\textrm{L}}(\varrho)=\sqrt{c^{2}-c_{\min }^{2}} \quad\text {and}\quad F^{(3)}_{\textrm{L}}(\varrho)=c,
\end{equation}
where $c=\sqrt{\vec{c}^{\,2}}$ and $c_{\min } \equiv \min \left\{\left|c_{1}\right|,\left|c_{2}\right|,\left|c_{3}\right|\right\}$.

Then, in this case, we have the linear steering measure~\cite{Costa2016} given by
\begin{equation}\label{lsm2}
S^{(2)}_{\mathrm{L}}(\varrho):=\max \left[0, \frac{\sqrt{c^{2}-c_{\min }^{2}}-1}{\sqrt{2}-1}\right],
\end{equation}
in the case of two measurement settings, and
\begin{equation}\label{lsm3}
S^{(3)}_{\mathrm{L}}(\varrho):=\max \left[0, \frac{c-1}{\sqrt{3}-1}\right],
\end{equation}
for three measurement settings.

\subsection{Generalized Entropic Steering Measure} 

After analyzing the linear steering criteria, we consider the entropic steering criteria presented in Equation~\eqref{tsc}. Optimizing the measurements in order to obtain $F^{(q,m)}_{\textrm{E}}(\varrho)$ is not trivial, even in the simple case of $q=2$. Although it is possible to perform a numerical optimization, we are interested in obtaining a closed analytical formula. Given this difficulty, we focus on the case of fixed orthogonal measurements based on the Pauli operators. This choice means that $\hat{u}_{i}=\hat{v}_{i}=\left\{(1,0,0)^{T},(0,1,0)^{T},(0,0,1)^{T}\right\}$ for $m=3$, and we also consider $q=2$, based on numerical investigations which show that this criterion is the strongest among this class. Thus, we obtain the following entropic steering measure (for $q=2$ and fixed measurements):
\begin{equation}\label{esm}
S^{(2,3)}_{\mathrm{E}}(\varrho):=\max \left\{0, 1-\sum_{r=1}^{3}\left[\frac{1-a_{r}^{2}-b_{r}^{2}-c_{r}^{2}+2 a_{r} b_{r} c_{r}}{2\left(1-a_{r}^{2}\right)}\right]\right\}.
\end{equation}

It remains an open question whether it is possible to obtain analytically a closed formula to quantify steering considering the criteria based on the Tsallis entropic uncertainty relations, for all values of $q$.

\subsection{Rotationally Invariant Steering Measure}

The rotationally invariant steering criterion~\eqref{SRI} is invariant under rotations {for a large class of states~\cite{Wollmann2016}}. Then, in this case, it is not necessary to perform an optimization over all measurements, as fixing Alice and Bob's measurements to the ones based on the Pauli operators is sufficient. In this case, we arrive at the following measure for a system of two~qubits
\begin{equation}\label{ris3}
S^{(3)}_{\mathrm{RI}}(\varrho):=\max \left[0,\frac{(|c_1|+|c_2|+|c_3|)-\sqrt{3}}{3-\sqrt{3}} \right].
\end{equation}

\subsection{Dimension-Bounded Steering Measure} 

The dimension-bounded steering criterion is very useful in detecting steering, given that one can simply assume that Bob's measurements act on a qubit system instead of trusting his devices. It is also rotationally invariant~\cite{Wollmann2020}. In order to obtain a steering quantifier for the dimension-bounded steering, one has to analyze the elements of the data matrix $\mathcal{D}$ constructed with mean values $\langle A_i \otimes B_j \rangle$, $\langle A_i \rangle$ and $\langle B_i\rangle$. To maximize $|\det\mathcal{D}|$ in order to obtain $F^{(m)}_{\textrm{DB}}(\varrho)$, we adopt a similar analysis to the linear case~\eqref{CJWR}. The determinant is optimal when the elements $\langle A_i \otimes B_j\rangle$, for $i\neq j$, are null, a condition for which each of Alice and Bob's measurement settings must be orthogonal, as we can verify directly from Equation~\eqref{ABmeanval}. Thus, the dimension-bounded steering measure is given by
\begin{equation}
\small 
S^{(3)}_{\textrm{DB}}(\varrho) = \max \left[0,\frac{3\sqrt{3}|c_1c_2c_3-(a_1b_1 c_2 c_3 + a_2b_2 c_1 c_3 + a_3b_3 c_1 c_2)|-1}{3\sqrt{3}-1} \right].
\end{equation}

It is interesting to note that we would arrive at the same result if we had chosen fixed measurements based on the Pauli operators. This result shows that the rotational invariance, first noted in~\cite{Wollmann2020}, is confirmed for this criterion.

\section{Steering Measures Comparison}

\begin{figure}[H]
{\includegraphics[height=1.9in]{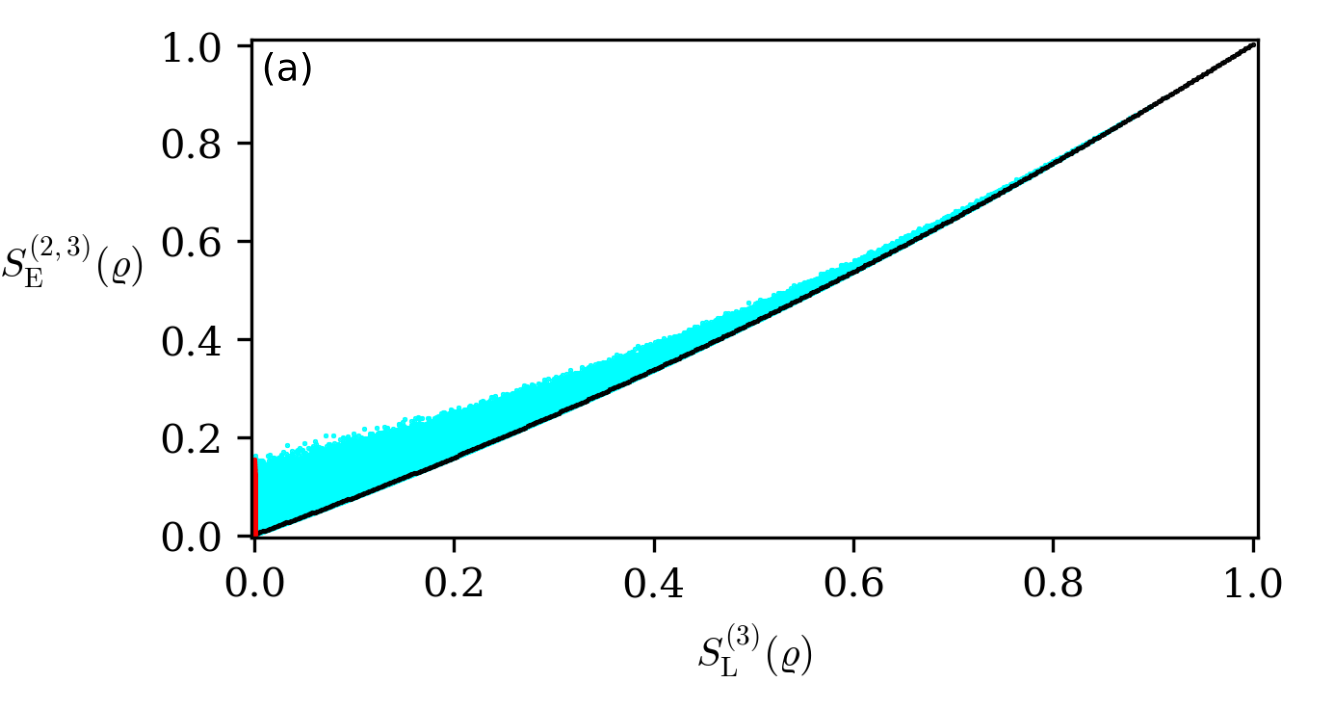}}\\
{\includegraphics[height=1.9in]{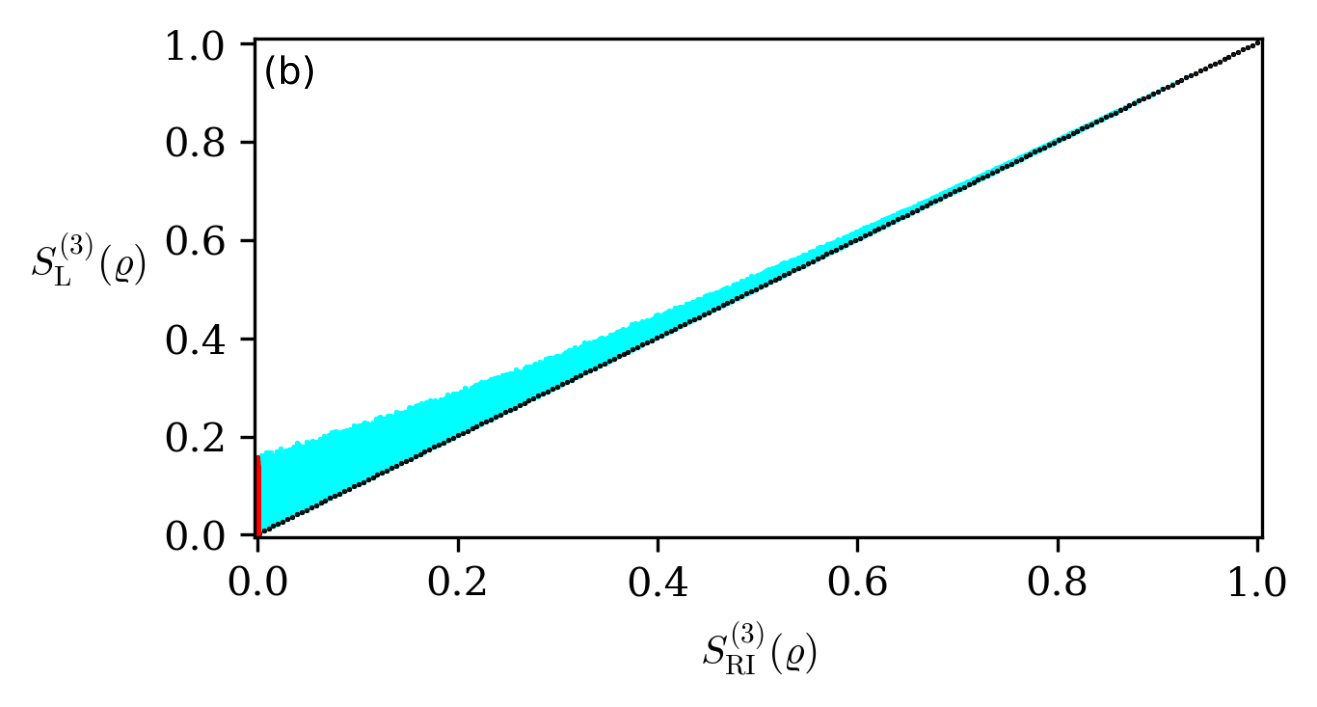}}\\
{\includegraphics[height=1.9in]{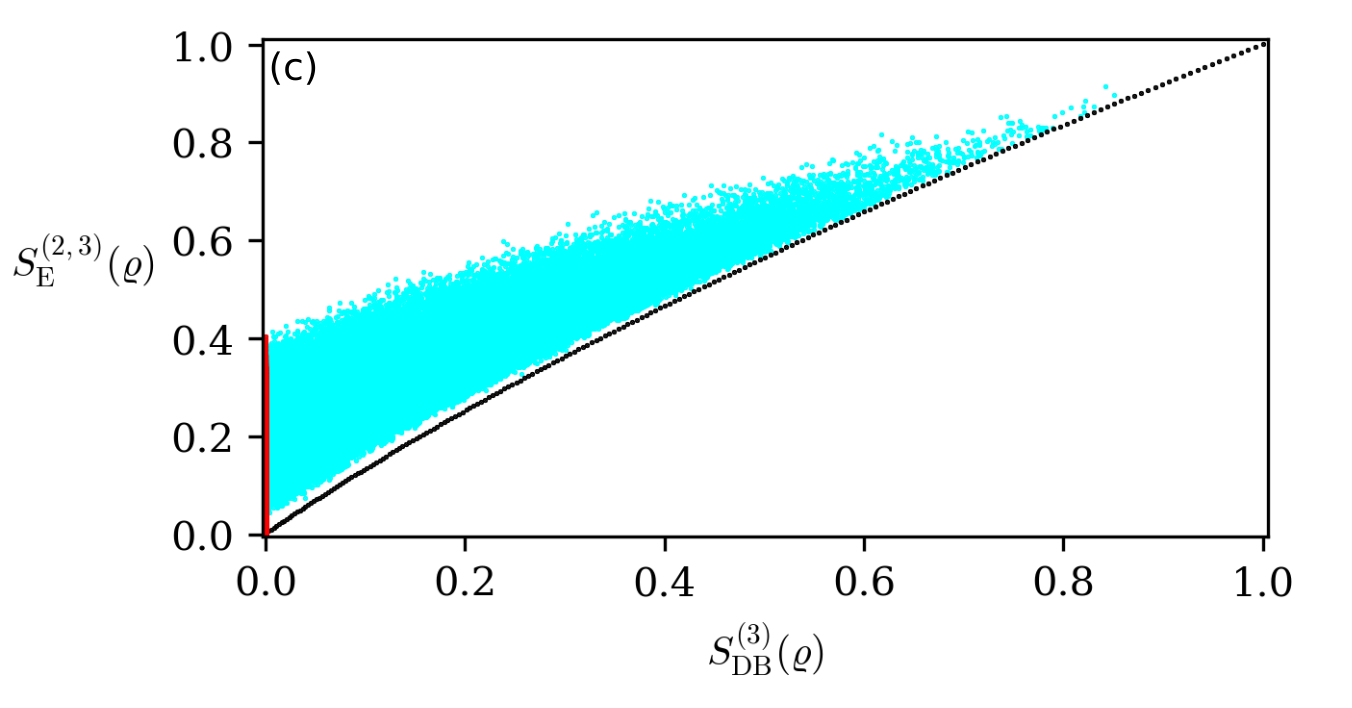}}
\caption{{Diagrams} 
of ({\bf a}) $S^{(2,3)}_E(\varrho)$ with $S^{(3)}_L(\varrho)$, ({\bf b}) $S^{(3)}_L(\varrho)$ with $S^{(3)}_{RI}(\varrho)$, and ({\bf c}) $S^{(3)}_{L}(\varrho)$ with $S^{(3)}_{DB}(\varrho)$. The cyan points correspond to $10^6$ general two-qubit states randomly generated. The red points represent when one of the measures is zero while the other one is not. The black points represent Werner states $\varrho_w$ with $w \in [0,1]$. The value $w=0$ ($w=1$) corresponds to the minimum (maximum) for all quantifiers.}
\label{phase}
\end{figure}

In the previous section, we presented steering quantifiers based on different criteria. These analytical formulas to quantify steering is one of our main results. Now, we can check if they are all equivalent, and if they are not, which one can detect more steerable states than the others. For this analysis, we generated $10^6$ random states and constructed parametric plots to compare the different quantifiers. In Figure~\ref{phase}, each point represents a random state while the black points represent the Werner states $\varrho_w = w \ketbra{\psi_s} + \frac{1-w}{4} \mathbbm{1}_4$, where $\ket{\psi_s}=(\ket{01}-\ket{10})/\sqrt{2}$ is the singlet state. For $w=0$, corresponding to maximally mixed states, all measures are null, and for $w=1$, being the maximally entangled states, all the quantifiers are maximal. In Figure~\ref{phase}, one can notice that the Werner states form a lower bound in the measures' comparison.

An important result, extracted from Figure~\ref{phase}, is the fact that the measure based on the generalized entropic direction~\eqref{esm} appears to be the most {sensitive} to quantify steering for two-qubit systems. When analyzing all the figures, it is possible to notice that some quantifiers disappear for certain states while the entropic one does not. {The fact that the entropic measure detects steering for a larger class of states, compared to the other quantifiers, makes it more efficient than the others.} It is our second important result, an analytical formula that appears to be more {resilient} to detect (and quantify) when compared to other steering criteria in the literature. {We emphasize that this} conclusion is only valid for the case of three measurements per set ({{this conclusion} 
can also be made for two measurements per set, but since taking three measurements provides stronger steering criteria, we restrict the analysis to the latter case}). Moreover, if we consider the possibility of carrying out an infinite number of measurements, better quantifiers can be found in~\cite{Jevtic2015,Nguyen2016}.

{Furthermore, it is important to note that although the measures are zero, this is not enough to demonstrate that the state is nonsteerable, {i.e.,} if the inequalities $F^{(m)}_{\mathrm{K}}(\rho,\nu) \leqslant \mathcal{B}^{(m)}_{\mathrm{K}}$ are not violated, it is also not sufficient to prove unsteerability.}


\section{Steering Robustness}

In the section above, we defined different steering measures based on the maximal violation of a respective steering inequality. However, it is important to note that there is also a quantifier known as steering robustness, based on numerical optimization, and structured on the framework of a semidefinite program (SDP) problem. Here, we present an overview of the SDP structure for steering (for more details on how to implement the algorithm, see~\cite{Cavalcanti2017}).

For a finite number of measurement choices and outcomes, the probability $p(a | x, \lambda)$ can be decomposed as a convex combination of a finite number of deterministic probability~\mbox{distributions}:
\begin{equation}\label{Dprob}
p(a|x, \lambda)=\sum_{\lambda^{\prime}} p\left(\lambda^{\prime} |\lambda\right) D\left(a | x, \lambda^{\prime}\right),
\end{equation}
where $p\left(\lambda^{\prime} | \lambda\right)$ is the weight of the deterministic distribution of the hidden variable $\lambda^{\prime}$ given $\lambda$. An LHS assemblage can be written by inserting (\ref{Dprob}) into (\ref{LHSassemblage}) in order to obtain the members of the assemblage as
\begin{equation}\label{sigma_det}
\sigma_{a | x}=\sum_{\lambda^{\prime}} D\left(a | x, \lambda^{\prime}\right) \sigma_{\lambda^{\prime}},
\end{equation}
where we have defined $\sigma_{\lambda^{\prime}}:=\int \mathrm{d} \lambda \mu(\lambda) p(\lambda^{\prime} |\lambda) \rho_{\lambda}$. Given that the distribution $D\left(a | x, \lambda^{\prime}\right)$ is fixed, in contrast with $p(a | x, \lambda)$ which was unknown, the model (\ref{sigma_det}) is a simplification of the LHS assemblage structure.

We are now able to implement an SDP to verify whether an assemblage $\left\{\sigma_{a | x}\right\}_{a, x}$ is a member of the LHS set:
\begin{equation}
\begin{array}{lll}
\text { given } & \left\{\sigma_{a | x}\right\}_{a, x} \; ;\{D(a | x, \lambda)\}_{\lambda} & \\
\text { find } & \left\{\sigma_{\lambda}\right\}_{\lambda} & \\
\text { s.t. } & \sum_{\lambda} D(a | x, \lambda) \sigma_{\lambda}=\sigma_{a | x} \quad \forall a, x. \\
& \sigma_{\lambda} \geqslant 0 \quad \forall \lambda .
\end{array}
\end{equation}
To construct a steering measure, we start by asking how much noise ($\varepsilon$) one has to add to a given assemblage in order for it to have an LHS model. For any convex subset of assemblages ($\mathcal{R}$) characterized by positive semidefinite constraints and linear matrix inequalities, we can define the steering robustness of an assemblage as
\begin{equation}\label{SR}
\begin{aligned}
\operatorname{SRA}\left(\{\sigma_{a | x}\}\right)= &\min_{\pi_{a | x}, \sigma_{\lambda}, \varepsilon}  \;\varepsilon \\
\text{s.t.} & \;\;\; \frac{\sigma_{a | x}+\varepsilon\, \pi_{a | x}}{1+\varepsilon}=\sigma_{a | x}^{\mathrm{LHS}} \quad \forall a, x, \\
& \;\;\; \sigma_{a | x}^{\mathrm{LHS}}=\sum_{\lambda} D(a | x, \lambda) \sigma_{\lambda} \quad \forall a, x, \\
& \;\;\; \pi_{a | x} \in \mathcal{R}, \quad \sigma_{\lambda} \geq 0 \quad \forall \lambda.
\end{aligned}
\end{equation}
The types of noise that can be applied and their respective robustness quantifiers are given from the subset $\mathcal{R}$. The next steps are to transform~\eqref{SR} explicitly into an SDP and then into an explicitly feasible problem. Details can be found in \cite{Cavalcanti2017}.

In the strategy used to derive the steering robustness quantifier, we need to fix a choice of measurements for Alice. Another approach would be to optimize over the possible choices of $M^A_{a|x}$, a method known as the see-saw algorithm. Formally, it can be presented as
\begin{equation}\label{srs}
\begin{aligned}
\mathrm{SRS}(\rho)= & \max _{\left\{M^A_{a | x}\right\}} \mathrm{SRA}(\{\sigma_{a | x}\}) \\
\text{s.t.}& \;\;\;\sigma_{a | x}= \Tr_{\mathrm{A}}\left[\left(M_{a | x} \otimes \mathbbm{1}\right) \rho\right] \quad \forall a, x.
\end{aligned}
\end{equation}
Using this construction, we would be able to find how steerable a given state is by the calculation of the steering robustness considering $m$ measurements and performing an optimization over all of them.

Since the see-saw algorithm requires a lot of computational effort (it incorporates a second maximization), we compared the statistics for a set of generic states $\varrho$ calculating their steerability via the steering robustness~\eqref{SR} assuming fixed measurements based on the Pauli operators and the steering robustness supported by the see-saw algorithm~\eqref{srs}. {As a result, in an analysis of $3 \times 10^4$ random states, for $12\%$ of the states, the latter presents higher values of steerability. Although it is not a low statistic, the difference between the values is around $10^{-3}$.} This result is not a surprise since the SDP in~\eqref{SR} performs an optimization over all possible inequalities to detect the steerability of a given state. Considering this, we focused on steering robustness with fixed measurements to perform the following analysis.

\begin{figure}[H]
\centering
\includegraphics[height=1.94in]{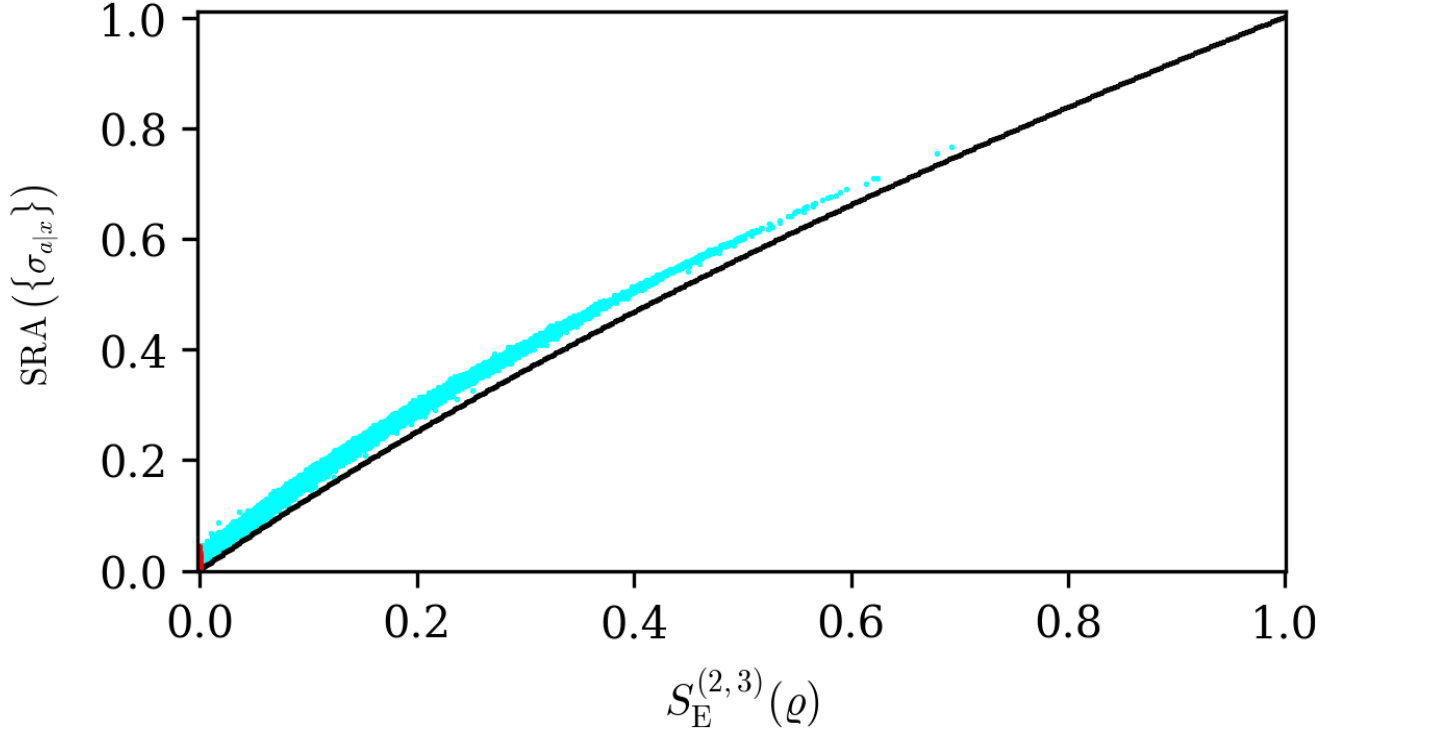}
\caption{Comparison of steering robustness $\operatorname{SRA}\left(\{\sigma_{a | x}\}\right)$ and the generalized entropic steering measure $S^{(2,3)}_{\textrm{E}}$ for $2.4\times 10^5$ random states. For the red points ($\thicksim$5.5\%), SRA is non-null while the entropic measure vanishes. The black points are the Werner states $\varrho_w$ with $w\in [0,1]$.}
\label{fig2}
\end{figure}

Finally, we compared the steering robustness (which is a numerical procedure to detect the steerability of a given state) with the closed formula provided by the Tsallis entropy based steering measure~\eqref{esm}, which we showed in the previous section to be the measure that best detected and quantified the steerability. For this analysis, we generated $2.4 \times 10^{5}$~random states and calculated both measures. The results can be visualized in Figure~\ref{fig2}. Given this set of states, around $5.5\%$ are only detected by the steering robustness and by the analytical formula provided in Equation~\eqref{esm}. This is one of the main results of this work. Although the closed formula to quantify steerability given in Equation~\eqref{esm} is not optimal, we still have a very good agreement with another quantifier that relies on a computational~method.

\section{Volume of Violations}

So far, we have compared the methods by generating a collection of states and calculating the value of the optimized quantifiers---or assuming orthogonal and aligned measurements between the parts---for each member of the set. Another way to compare different approaches to detect steering is by fixing a quantum state and performing on it a collection of measurements, which are orthogonal but randomly distributed. In the end, we obtain a statistic of how many rounds of measurements violated each quantifier.

This idea was proposed by Fonseca and Parisio~\cite{Fonseca2015} in order to quantify Bell's nonlocality, {motivated by the fact that almost all known Bell inequalities for more than two outcomes are maximally violated by states that are not maximally entangled~\cite{Acin2002}}. They defined the space of all possible parameters $\text{Z}=\{x_1, \ldots, x_m, y_1, \ldots, y_n\}$ that determined the settings for a Bell-like inequality. For a state $\rho$, the set of points $\zeta_\rho \subset \text{Z}$ that lead to a violation of the inequality will have a corresponding quantity $V(\rho)$ proportional to the volume of $\zeta_\rho$. The authors proposed that if $V(\rho)>V(\rho')$, then the state $\rho$ is more nonlocal than the other $\rho'$, with
\begin{equation}
V(\rho) := \frac{1}{\mathcal{N}} \int_{\zeta_{\rho}} d^{m} x d^{n} y,
\end{equation}
where $\mathcal{N}$ is a constant of normalization.

In the following analysis, we investigate the quantification of steering for all the criteria presented in Section~\ref{ineq} using the approach based on the volume of violations. This gives us another method to quantify the steerability of a given state, with the difference that here, we do not optimize over all possible measurements, but we check the volume of violations of a specific inequality. Unlike the previous method, here, it is not possible to derive analytical formulas.

\begin{figure*}[!htb]
    \centering
  {\includegraphics[height=1.95in]{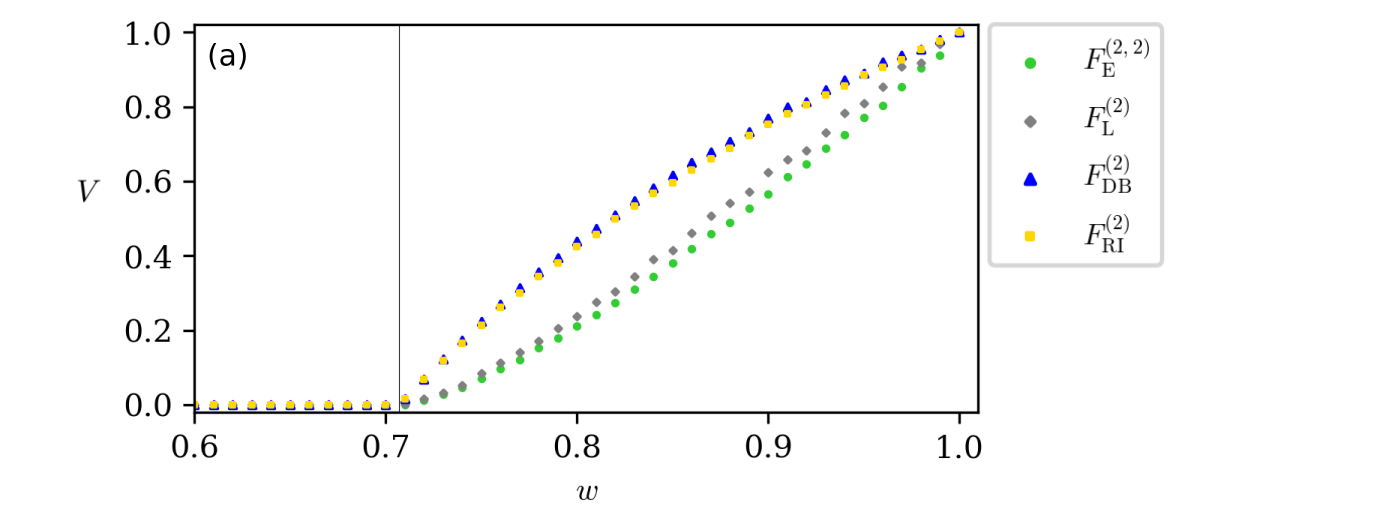}}
  {\includegraphics[height=1.95in]{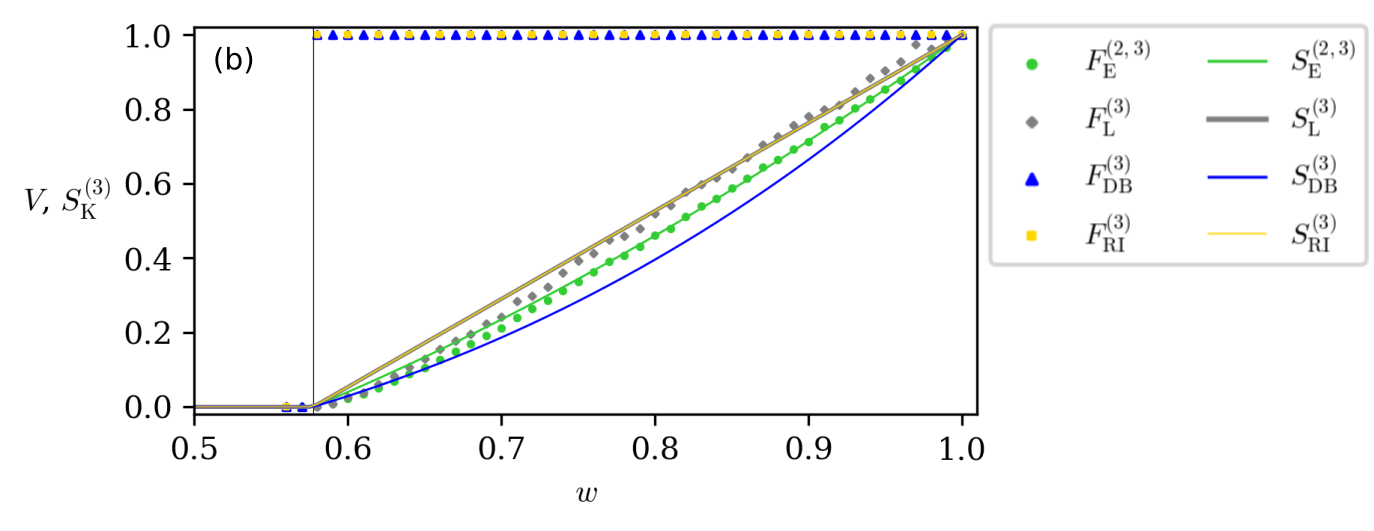}}
    \caption{Comparison between the different steering measures for ({\bf a}) $m=2$ and ({\bf b}) $m=3$  measurement settings per site. Each point represents a Werner state whose volume of violations was computed for $10^5$ measurements. In ({\bf b}) we also plotted the analytical formulas presented in Section~\ref{measures} to compare both ways of quantification.}
    \label{fig3}
\end{figure*}

Using the approach presented in~\cite{Fonseca2015}, we checked, for specific Werner states, the volume of violations for the inequalities of Section~\ref{ineq}. The results can be seen in Figure~\ref{fig3}, where we performed numerical simulations for $10^5$ orthogonal random measurements. When considering two measurement settings per site, one can see that there are no significant differences between the criteria. They all give approximated values, and, most importantly, they are all null and maximal for the same states, showing some notion of equivalence between them.

{For the case of three measurement settings per site,} the rotationally invariant steering~\eqref{SRI} and the dimension-bounded steering~\eqref{F_DB} give different results when compared to the linear steering~\eqref{CJWR} and the generalized entropic steering~\eqref{esm} for $q=2$. This result is not surprising, given that the former is known to be invariant under rotations so they would violate a larger number of measurements than the latter one. 
Interestingly, according to this quantification method, all Werner states with $w>1/\sqrt{3}$ are maximally steerable, according to the volume of violations of the dimension-bounded and the rotationally invariant~\mbox{steering}.

{Furthermore, it is interesting to note that the rotationally invariant and the dimension-bounded steering measures based on the volume of violations, unlike the others, do not show a convex behavior as the purity of the states increases. Although this is pointed out as an important property that a steering measure should satisfy, as discussed in~\cite{Gallego2015}, it is not mandatory. In this sense, they still can be used as quantum steering quantifiers, although they show an atypical behavior.}

To complement our analysis, we also plotted the analytical measures developed in Section~\ref{measures} in Figure~\ref{fig3}b. It is very interesting that except for the dimension-bounded and the rotationally invariant steering with three measurements per site, both ways of quantification have a similar behavior, the generalized entropic steering being the one that fits better with each other.
\vspace{-3pt}


\section{Discussion}
\label{Discussion} 

In this work, we analyzed two methods to quantify steerability, one based on the maximization over all measurements and the other based on the volume of violations of steering inequalities. Using the former method, we obtained analytical formulas that can be used to quantify steering. We focused on four known criteria from the literature, the linear steering~\cite{Cavalcanti2009}, the generalized entropic steering~\cite{Costa2018,Costa2018-2}, the rotationally invariant steering~\cite{Wollmann2016}, and the dimension-bounded steering~\cite{Moroder2016}. Among the quantifiers, the generalized entropic measure~\eqref{esm} was more efficient in detecting and quantifying steering for a larger class of states. In the sequel, we compared this measure to another well-known quantifier, the steering robustness, which is based on semidefinite programming. The computational method proved to be more efficient only for $5.5\%$ of the states.

While considering the volume of violations of these inequalities, the results showed that this method gave similar results to the one based on the maximization of the measurements, except for the rotationally invariant criteria. Here, we focus only on the family of Werner states, which already showed us interesting results.

For future work, it would be interesting to extend this analysis to higher dimensional states, where the numerical method is more demanding, and obtaining an analytical formula to quantify steering for this class of states could be very helpful for different applications. Another aspect to be investigated is the extension to multipartite systems, where it could be possible to look at the monogamy of steering. 

\vspace{6pt} 






\bibliography{references} 
\end{document}